Comments to "Comparison of two experiments on radiative neutron decay"
by R.U.Khafizov et al., published in *Physics of Atomic Nuclei*,
2009, Vol. 72, № 12, pp. 2039-2048.

I.A.Kuznetsov

Petersburg Nuclear Physics Institute RAS

188300, Gatchina, Russia

In the paper published in "Physics of Atomic Nuclei" ("Yadernaya Fizika") [1] authors drew a comparison between two experiments [2,3] on search of radiative neutron decay and measurement of its branching ratio BR.

This publication causes some bewilderment at least. Not covering here the experiment of NIST group [3], let's observe the result published in [2]. The first stage of this experiment was made at ILL high flux reactor (Grenoble) and the main statistics was collected at FRM-II reactor in Munich.

The spectrum of double **e-p** coincidences, i.e. proton time-of flight spectrum when signals from the electron detector used as "start", is shown in Fig.5. This spectrum contains a lot of peaks, and only two major of them are explained – the peak of prompt coincidences (in channel 99) and the peak of protons from neutron decay with maximum in channel 120. Other smaller peaks (authors said about 7 peaks) have no explanation. The words in [1] about problems with the electric circuit seem to be unconvincing. Such data should be rejected in a real precise experiment.

The spectrum of triple **e-p-γ** coincidences between three different detectors is shown in Fig.6. There are no clear explanations in the figure caption what time represents on the horizontal axis. It was said in the text that the feature in "channel 120 in Fig.6" contains coincidences between decay electrons and protons, along with an event in one of the gamma detectors. And this spectrum looks a little like spectrum of double **e-p** coincidences in Fig.5, but rather distorting. But the authors have approved the spectrum in Fig.6 is the spectrum of **e-γ** coincidences under a condition when a signal from the electron detector coincides with a signal from the proton detector. Also they approved that "the leftmost peak in channel 103 is connected to the in question as this gamma-quantum has been registered in our equipment before the electron".

It should be mentioned that the distance between decay area and an electron (or gamma detectors) is about 30 cm. It means that electrons (or gammas) pass through this 30 cm for 1-2 nanoseconds. Such conditions make it impossible to resolve "the peak of the radiative gamma-quanta" and the peak of prompt coincidences taking into account the width of the time channel 25 ns.

On the other hand, following the formal logic of the authors, let's look at formula (1) in Ref.[2] accurately

$$S_{out}(t) = \int S_{in}(t') R_\gamma(t, t')dt'.$$

Then the authors wrote: "In our case $S_{in}(t) = S_D(t)$ is the spectrum of double **e-p** coincidences in Fig.5, while $S_{out}(t)$ is the background spectrum of the triple coincidences." Moreover, the authors introduced the response function of gamma channel $R_\gamma(t, t')$ as for the situation with perfect detector $R_\gamma(t,t') = k\delta(t- t')$, i.e. δ-function with coefficient k which defined by the total gamma - detector's count rate. But such an approach has no relation to the radiation neutron decay. It means only the transposition of spectrum of double **e-p** coincidences onto the gamma-event scale. And such transposition makes literally "channel to channel". In other words, for every double coincidence in some channel (not only in the proton peak region) in Fig.5 the authors search the signal on the gamma-scale in the same channel. It means that the time window on gamma scale is very narrow, about 1 channel or 25 ns. The discrepancy between spectra of double and triple coincidences the authors explained as "for real detectors the response function is always not local which leads to a deformation of spectrum $S_{out}(t)$".

In such approach, it is physically impossible to select radiative neutron decay events. The left peak in channel 106, Fig.6 have no relation to the decay proton at all (to speak about "radiative gamma quanta peak in channel103" is not correct due to bad time resolution, see above). Of course, there are γ-quanta from radiative decay in this peak, but it's impossible to separate it from prompt events.

Also, it should be noted that these remarks are not the first critical comments of this experiment (see [4]).

Using such layout of the triple coincidence definition it's impossible to carry out an experiment correctly, and the value of BR obtained in this measurement couldn't be recognized as correct.

**References**

1. R.U. Khafizov *et al.*, Comparison of two experiments on radiative neutron decay. **Physics of Atomic Nuclei**, Vol.**72**, № 12, pp.2039-2048, (2009) ("Yadernaya Fizika", **72**, № 12, pp.2102-2111, (2009)).

2. R.U. Khafizov *et al.*, Observation of the neutron radiative decay. **JETP Let**t. **83**, 1, pp. 5–9 (2006).

3. Nico, J. S. *et al*., Observation of the radiative decay mode of the free neutron. **Nature**, **444**, pp.1059-1062 (2006).

4. Severijns, N., Zimmer, O., Wirth, H.-F.&Rich, D. Comment on ''Observation of the neutron radiative decay'' by R. U. Khafizov et al., **JETP Lett 84**, 231 (2006).